\documentclass[pra,reprint,superscriptaddress,longbibliography]{revtex4-1}
\usepackage{braket}
\usepackage{amsmath}
\usepackage{graphicx}
\usepackage[utf8]{inputenc}
\usepackage{natbib}
\usepackage[section]{placeins}
\usepackage{pgfplots,amsfonts,mathtools}
\usepackage[breaklinks]{hyperref}
\usepackage{amssymb}
\usepackage{amsmath,amsthm,bm}
\usepackage{amsfonts}
\usepackage{cleveref}
\usepackage{color,soul}
\DeclareMathOperator*{\argmin}{argmin}
\hypersetup{citecolor=red,colorlinks=true,urlcolor=blue}

\usepackage[cal=stixtwofancy]{mathalpha}

\begin{document}
\title{Simulating Quantum Many-Body States with Neural-Network Exponential Ansatz}

\author{Weillei Zeng}
\affiliation{Zhejiang Laboratory, Hangzhou 311100, China}

\author{Jiaji Zhang}
\affiliation{Zhejiang Laboratory, Hangzhou 311100, China}

\author{Lipeng Chen}
\email{chenlp@zhejianglab.com}
\affiliation{Zhejiang Laboratory, Hangzhou 311100, China}

\author{Carlos L. Benavides-Riveros}
\email{cl.benavidesriveros@unitn.it}
\affiliation{Pitaevskii BEC Center, CNR-INO and Dipartimento di Fisica, Università di Trento, I-38123 Trento, Italy}

\date{\today}

\begin{abstract}
Preparing quantum many-body states on classical or quantum devices is a very challenging task that requires accounting for exponentially large Hilbert spaces. Although this complexity can be managed with exponential ansätze (such as in the coupled-cluster method), these approaches are often tailored to specific systems, which limits their universality. Recent work has shown that the contracted Schrödinger equation enables the construction of universal, formally exact exponential ansätze for quantum many-body physics. However, while the ansatz is capable of resolving arbitrary quantum systems, it still requires a full calculation of its parameters whenever the underlying Hamiltonian changes, even slightly. Here, inspired by recent progress in operator learning, we develop a surrogate neural network solver that generates the exponential ansatz parameters using the Hamiltonian parameters as inputs, eliminating the need for repetitive computations. We illustrate the effectiveness of this approach by training neural networks of several quantum many-body systems, including the Fermi-Hubbard model.
\end{abstract}

\maketitle

\section{Introduction}

Rapid progress in quantum technologies promises substantial computational advances in a diverse range of fields, from simulating quantum many-body systems with quantum devices to developing hybrid quantum-classical algorithms for optimization tasks \cite{Lloyd96,RevModPhys.86.153,Acín_2018,Bauer2020,sommaQuantumSimulationsPhysics2003,doi:10.1126/sciadv.adm6761}. Among the numerous challenges ahead, a critical aspect of simulating the dynamics of quantum many-body systems on quantum hardware is the efficient preparation of quantum states described in exponentially large Hilbert spaces. Although such quantum states can be prepared using, for example, adiabatic techniques \cite{Veis2014,PhysRevLett.104.030502} or quantum imaginary time evolution \cite{Motta2020,doi:10.1021/acs.jctc.3c00071}, these methods require selecting non-degenerated evolution paths or performing time propagation (either imaginary or real), which, depending on the Hamiltonian spectrum or the initial overlap with the target state, can be highly time-consuming, computationally demanding, or even unfeasible in practical situations  \cite{Bauer2020}. 

An alternative approach to quantum state preparation, which is also applicable to estimating eigenstates on classical computers, leverages well-designed ansätze that reduces the complexity of the wave function while retaining its essential structural features. In quantum chemistry, for example, a powerful class of exponential wave-function ansätze is available within the coupled-cluster theory, both in its standard non-unitary form \cite{JCizek_1980, RevModPhys.79.291} and in its unitary variant \cite{doi:10.1021/acs.jctc.8b01004,D1CS00932J,Shavitt_Bartlett_2009}. When impurities are present, however, quantum systems often undergo significant changes that limit the effectiveness of simpler perturbative or mean-field approaches \cite{grusdt2015new}. Consequently, various ansätze have been proposed to address these complexities, including the Chevy (or polaron) ansatz for polarized atomic Fermi gases at zero temperature \cite{PhysRevA.74.063628,grusdt2024impuritiespolaronsbosonicquantum}, variational ansätze for dressed dimers \cite{Combescot_2009}, or more sophisticated ansätze for rotating molecules in bosonic environments \cite{doi:10.1080/00268976.2019.1567852}.

Although the need to engineer wave-function ansätze spands across various domains of quantum physics, no universal framework has yet emerged for constructing ansatz of general quantum many-body systems \cite{szenes2024striking, Magoulas2023}. In fact, each field of research (and, in some cases, each interaction range within a research field) tends to develop its own set of wave-function ansätze, resulting in a fragmentation of results that poses significant challenges for drawing broad conclusions about different phases of matter, transferring knowledge between sub-fields of research, or studying complex phenomena involving various types of excitations (e.g., fermionic or bosonic). An even more significant problem is that parameterizing such wave-function ansätze for different Hamiltonian parameters (e.g., the classical nuclear coordinates in molecular systems) requires high-dimensional configuration spaces. As a result, there is a growing need for an efficient, formally universal, ab initio framework capable of designing quantum many-body ansätze, one that does not rely on the specific form of the underlying Hamiltonian and can potentially be implemented in near-term quantum simulators \cite{doi:10.1021/acs.jpclett.1c02659,f.ribeiroPolaritonChemistryControlling2018,duCatalysisDarkStates2022,mordovinaPolaritonicCoupledclusterTheory2020,doi:10.1021/acs.jctc.8b00932}.

Recent progress in formulating the quantum many-body problem using the contracted Schrödinger equation has highlighted the existence of a universal, formally exact exponential ansatz \cite{Mazziotti.20060v3}. This is an iterative ansatz that can be used to prepare ground and excited states for electronic and bosonic systems, as well as for fermion-boson or boson-boson mixtures, such as those encountered in polaritonic quantum chemistry or supersolidity \cite{PhysRevLett.126.070504, smart2023verifiably,benavidesriveros2023quantum,Smart.2022,warren2024exact,PhysRevResearch.6.L012052}. Beyond this remarkable universality, a key feature of this ansatz is that it retains precisely the same number of degrees of freedom as the original quantum many-body Hamiltonian. Unlike other exponential ansätze, the number and the operational form of the terms in the exponent are fixed and correspond to those of the Hamiltonian \cite{Boyn2021a, 10.1063/5.0074842}. Yet, despite this remarkable mathematical structure, the family of contracted quantum (and classical) eigensolvers proposed in the literature requires an indefinite number of iterative calculations, along with a complete recalculation of its parameters whenever the underlying Hamiltonian changes, even slightly. This limitation diminishes the universality of the ansatz and constrain its capacity to explore diverse correlation regimes or uncover new states of matter. 

In this paper, inspired by recent advances in operator learning, we develop a surrogate neural network solver that directly outputs the parameters of the universal, formally exact exponential ansatz of quantum many-body physics. By design, our neural network captures the implicit nonlinear functional relationship between the Ha\-mil\-tonian and the ansatz parameters. This approach not only eliminates the need for repetitive and computationally demanding calculations but also reinforces the method's universality. 

The paper is organized as follows. In the first part, we discuss the main ingredients of the exact exponential ansatz and present its main connections with the family of contracted quantum eigensolvers. We introduce a crucial modification to those eigensolvers that makes the learning process unambiguous. Next, we introduce our surrogate neural solver for the exact ansatz and present our numerical results. We finish with a conclusion section where we also discuss potential implications of our results.

\section{Theory} 

In this section, we discuss the general scheme of the universal, formally exact exponential ansatz of quantum many-body physics and its relation with the family of contracted quantum eigensolvers. We modify the contracted quantum eigensolver to make it suitable for learning tasks and, finally, we present our surrogate model. 

\subsection{Exact exponential ansatz for quantum many-body systems}

Let's start by writing a general many-body $M$-term Ha\-mil\-tonian as:
\begin{align}
    \hat H(\bm{f}) &=  \sum^M_{\mathcal{l}=1} f_\mathcal{l} \hat{h}_\mathcal{l},\label{Hamiltonian}
\end{align}
where $\hat{h}_\mathcal{l}$ is a general many-body operator, comprising the creation and annihilation of the corresponding (fermionic, bosonic, spin) modes and $\bm{f} \equiv \{f_\mathcal{l}\}$ is the set of $M$ Hamiltonian parameters. For instance, in the case of electronic systems with pairwise interaction $\mathcal{l} \equiv (ijkl)$ and thus 
\begin{align}
\hat h_\mathcal{l} = \hat c_i^\dagger \hat c_j^\dagger \hat c_k \hat c_l.
\end{align}
The Hamiltonian's parameters $f_\mathcal{l}$ correspond to the known 1- and 2-particle integrals. The stationary quantum many-body problem relies on the solution of the well-known Schr\"odinger equation:
\begin{equation}
\label{se}
    \hat{H}(\bm{f})\ket{\Psi_m(\bm{f})} = E_m(\bm{f}) \ket{\Psi_m(\bm{f})}.
\end{equation}
To obtain the so-called contracted Schr\"odinger equation (CSE) one multiplies this equation on the left by $\bra{\Psi_m(\bm{f})}\hat{h}_\mathcal{l}$ to obtain \cite{mazziottiContractedSchrodingerEquation1998,Valdemoro2007,PhysRevA.57.4219}: 
\begin{equation}
    \bra{\Psi_m(\bm{f})}\hat{h}_\mathcal{l}
    \hat{H}(\bm{f})\ket{\Psi_m(\bm{f})} = E_m(\bm{f}) \, D_{\mathcal{l},m}(\bm{f}),
    \label{eq:NCSE}
\end{equation}
where 
$D_{\mathcal{l},m}(\bm{f}) \equiv  \bra{\Psi_m(\bm{f})}\hat{h}_\mathcal{l}
    \ket{\Psi_m(\bm{f})}$ 
is the expectation value of the operator $\hat h_\mathcal{l}$. It is straightforward to see that this latter term can be related to certain terms of the corresponding multi-particle reduced density matrices of the $m$-th quantum state. For the case of electronic or bosonic systems that interact pairwise, this is the 2-body reduced density matrix \cite{RevModPhys.35.668,PhysRevLett.108.263002,PhysRevLett.124.180603}. Hybrid electron-boson mixtures will lead to richer reduced-density matrices. 

As already stated several times in the literature, the equivalence of Eq.~\eqref{eq:NCSE} and Eq.~\eqref{se} can be readily proved by multiplying both sides of the generalized CSE by the Ha\-mil\-tonian's parameter $f_\mathcal{l}$. Summing the expression over the indices $\mathcal{l}$ yields the well-known energy variance equation
\begin{align}
\bra{\Psi_m(\bm{f})} \hat{H}^2(\bm{f})\ket{\Psi_m(\bm{f})} = \bra{\Psi_m(\bm{f})} \hat{H}(\bm{f})\ket{\Psi_m(\bm{f})}^2, \nonumber
\end{align}
which, as a stationary condition for the set of eigenfunctions $\ket{\Psi_m(\bm{f})}$, is an equivalent formulation to the Schr\"o\-dinger equation \cite{mazziottiQuantumChemistryWave2006, Mazziotti.20060v3, Valdemoro2007}. As a result, the set of solutions to the CSE \eqref{eq:NCSE} must be the same as the solutions to the standard many-body Schr\"odinger equation \eqref{se}. Although this derivation is fully general and well known in the context of quantum chemistry as Nakatsuji's theorem \cite{nakatsujiEquationDirectDetermination1976,mazziottiContractedSchrodingerEquation1998,Valdemoro2007,PhysRevLett.93.030403,10.1063/1.1390516,MUKHERJEE2004174, PhysRevA.69.012507,PhysRevLett.91.123002}, it has been barely used beyond the realm of the electronic structure theory. Only recently, the CSE was used to compute ground-state properties of bosonic \cite{wang2023boson} and polaritonic quantum chemical systems \cite{warren2024exact}, as well as for electronic excited states \cite{benavidesriveros2023quantum,Wang.20232b}. 

It can be shown that the exact eigenstates satisfying the CSE  \eqref{eq:NCSE} can be found by implementing the following iterative ansatz \cite{PhysRevA.69.012507}:
\begin{align}
\ket{\Psi_m^{(n)}(\bm{f})} = e^{\hat B^{(n)}(\bm{f})} e^{\hat{A}^{(n)}(\bm{f})} \ket{\Psi^{(n-1)}_m(\bm{f})},
\label{ansatz0}
\end{align}
where  
\begin{align}
\hat A^{(n)}(\bm{f}) = \sum_{\mathcal{l}=1}^M A^{(n)}_{\mathcal{l}}(\bm{f})\,\hat h_\mathcal{l}
\end{align}
and 
\begin{align}
\hat B^{(n)}(\bm{f}) = \sum_{\mathcal{l}=1}^M B^{(n)}_{\mathcal{l}}(\bm{f})\, \hat h_\mathcal{l}
\end{align}
are, respectively,  anti-Hermitian and Hermitian operators, making thus the exponential ansatz a sequence of unitary and non-unitary operators acting on a given trial wave function. Notice that the ansatz \eqref{ansatz0} has the key property that the number of terms of each exponential operator is the same as the number of degrees of freedom of the original Hamiltonian, which ensures its scalability. 

The flexibility of this ansatz deserves to be highlighted. In particular, notice that if one fixes the non-unitary term as $B^{(n)}_{\mathcal{l}}(\bm{f}) = f_{\mathcal{l}}$, $\forall n$ (and the contributions of $\hat A^{(n)}(\bm{f})$ are suppressed) the ansatz coincides with the quantum imaginary time evolution technique which is known to give the exact ground state for $n\rightarrow \infty$ \cite{Motta2020,McArdle.2019}. Restricted to be unitary, the ansatz \eqref{ansatz0} generalizes, for electronic structure theory, the unitary coupled-cluster (UCC) ansatz \cite{doi:10.1021/acs.jctc.8b01004, Mazziotti.2007k2h}, whose generalization rediscovers indeed the unitary part of the ansatz, and has a strong connection with well-known canonical transformations of Hamiltonians involving fermion-boson coupling \cite{warren2024exact}.  As described in several previous works for the electronic-structure problem, this iterative ansatz can be used to converge to stationary states either through classical \cite{Mazziotti.20060v3, Gidofalvi.2009, Alcoba.2011} or quantum \cite{PhysRevLett.126.070504, Smart.2024, Boyn.2021u94, Smart.2022w8u, wang2023boson,benavidesriveros2023quantum} computing methods.

\subsection{Contracted quantum eigensolver with $k$ iterations}

In the literature, exponential ans\"atze can improve their computational efficiency by using circuit- and parameter-efficient iterative methodologies to select pools of cluster operators \cite{Fan2023}
or by closing the corresponding Lie algebra \cite{D0CP01707H}. Yet, in both cases, it is difficult to avoid an al\-ge\-braic explosion of the operator pooling. This is in stark contrast to the CSE ansatz \eqref{ansatz0}, which, as already mentioned, employs a set of operators whose size is always the same as the Hamiltonian degrees of freedom. However, while the ansatz is exact (in the sense that the convergence is guaranteed), it is unclear how many iterations are needed to obtain, within a certain energy (or fidelity) threshold, a good approximation for the desired quantum state. 

By focusing on the ground-state problem, we implement in this work a version of the contracted quantum eigen\-solver (that can be traced back to Ref.~\cite{PhysRevA.69.012507}) where the full minimization of the variational parameters is performed at each iteration as follows: 
\begin{align}
    \{A^{(n)}_{\mathcal{l}}(\bm{f}),B^{(n)}_{\mathcal{l}}(\bm{f})\} = \argmin_{\{A^{(n)}_{\mathcal{l}},B^{(n)}_{\mathcal{l}}\}} \bra{\Phi^{(n)}} \hat H (\bm{f})  \ket{\Phi^{(n)}},
    \label{min}
\end{align}
where 
\begin{align}
    \ket{\Phi^{(n)}} = e^{\sum_{\mathcal{l}=1}^M B^{(n)}_{\mathcal{l}}\hat h_\mathcal{l}}e^{\sum_{\mathcal{l}=1}^M A^{(n)}_{\mathcal{l}}\hat h_\mathcal{l}}\ket{\Psi^{(n-1)}(\bm{f})}\,.
\end{align} 
The quantum states $\ket{\Psi^{(n)}(\bm{f})}$, which should be normalized, are defined as in Eq.~\eqref{ansatz0}. On modern quantum devices, the minimization \eqref{min} can be performed by measuring the total residual of CSE, which can then be used to guide trial wave functions toward eigenstates by iteratively applying a sequence of exponential transformations \cite{Smart.2024, Boyn.2021u94, Smart.2022w8u, Wang.20232b, wang2023boson}.  

Finally, inspired by the $k$-UCC ansatz that takes a product of $k$ unitary operators to increase the ﬂexibility of the wave function \cite{doi:10.1021/acs.jctc.8b01004}, we restrict the number of (unitary and non-unitary) operators in Eq.~\eqref{min} to a maximum number $k$ such that the sought-after quantum state can be written as:
\begin{align}
\ket{\Psi(\bm{f})} = e^{\hat B^{(k)}(\bm{f})} e^{\hat{A}^{(k)}(\bm{f})} \cdots e^{\hat B^{(1)}(\bm{f})} e^{\hat{A}^{(1)}(\bm{f})} \ket{\Phi},
\label{ansatzfinal}
\end{align}
where $\ket{\Phi}$ is a trial wave function, that can eventually depend also on $\bm{f}$ if, e.g., a mean-field approximation is used. We call this approach $k$ contracted quantum eigensolver ($k$-CQE). So far we have focused on the many-body ground-state problem, but the generalization to excited states is straightforward if one makes use of the Rayleigh-Ritz variational principle for ensembles of quantum states \cite{benavidesriveros2023quantum,PhysRevLett.129.066401,Castillo2023,PhysRevLett.130.106401}.

\subsection{Neural network exponential ansatz} 

The possibility of using quantum data to learn the map between the parameters of the Hamiltonian and the parameters of exponential ans\"azte has been already suggested in several prior works \cite{ceroni2023generating, PRXQuantum.2.020329, nakaji2024generative, Lewis2024}. While their idea is in principle general, due to the use of the UCC ansatz for electronic systems, it has been only tested for molecular ground states. A major drawback of this type of approach, however, is the intrinsically non-universal form of the ansatz which includes an indeterminate number of variational parameters that typically depend on the dimensionality of the chosen active space or the structure of the excitation operators. In contrast, the ansatz \eqref{min} is universal (i.e., not restricted to the electronic structure theory) and the number of terms in the exponent is exactly equivalent to degrees of freedom of the Hamiltonian. 

Beyond quantum computing, the use of neural networks as surrogate solvers for specific types of differential equations has become an active field of research \cite{kovachki2021fno, Brunton2024, Xue2022}. Although the concept is generally applicable, discussions often focus on differential equations, as the solver can be formulated explicitly in terms of variables like initial and boundary conditions and corresponding solutions. Recently, a variety of Fourier-trans\-form-based architectures has been developed, showcasing a more flexible approach with applications to various classical dynamical problems \cite{alkin2024upt, guibas2021afno}, including applications in quantum sciences to scattering and quantum dynamics \cite{doi:10.1021/acs.jpclett.4c00598, PhysRevD.108.L101701,zhang2024neuralquantumpropagatorsdrivendissipative}.

\begin{figure}[t]
\includegraphics[width = \linewidth]{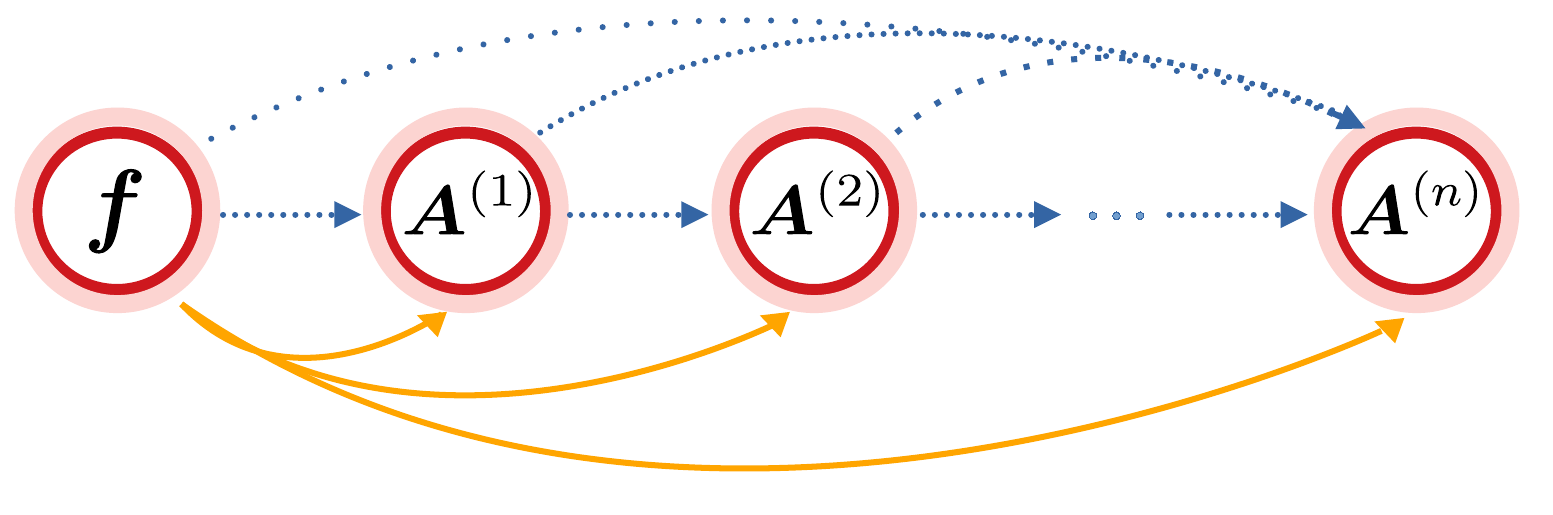}
\caption{Sketch of the learning schemes of the unitary part of the exact many-body ansatz. While the contracted quantum eigensolver learns each set $\bm{A}^{(n)}$ sequentially from the previous ansatz parameters $\{\bm{A}^{(m)}\}^{n-1}_{m=1}$ and the Hamiltonian parameters $\bm{f}$, the surrogate neural solver introduced in this work learns all $\{\bm{A}^{(m)}\}^{n}_{m=1}$ from $\bm{f}$. The former is illustrated in dotted blue lines and the latter in orange solid lines.}
\label{fig:learning}
\end{figure}

For our eigenstate problem, the explicit form of the functional relationship between the Hamiltonian parameters (i.e., ${\bm{f}}$) and the wavefunction ansatz parameters, (i.e., $\{A_{\mathcal{l}}(\bm{f}), B_{\mathcal{l}}(\bm{f}) \}$) is unknown, leaving it uncertain whether it can be learned. In fact, while Eq.~\eqref{min} suggests a sophisticated relationship between ${\bm{f}}$ and all those ansatz parameters, the complex nonlinear recursion between the Hamiltonian and the ansatz parameters, where each set is linked to the previous ones, prevents a clear, compact equation. Neural networks, however, are known for their capacity to approximate implicit functions, as demonstrated in applications such as neural machine translation \cite{Koehn_2020}. This makes neural networks a natural choice as surrogate eigensolvers for this complex, counterintuitive functional form.

In this work, we advance the construction of the implicit functional relationship between the Hamiltonian and the ansatz parameters.
Specifically, given a parameterized Hamiltonian $\hat H(\bm{f})$, the objective is to train a neural network as a surrogate model $\mathcal{G}_{\theta}$ that learns the following mapping:
\begin{align}
\{\bm{A}^{(n)}, \bm{B}^{(n)}\}_{n=1}^{k} = \mathcal{G}_{\theta}[\bm{f}] \,,
\end{align}
where $\bm{A}^{(n)} \equiv \{A^{(n)}_{\mathcal{l}}(\bm{f})\}$ and $\bm{B}^{(n)} \equiv \{B^{(n)}_{\mathcal{l}}(\bm{f})\}$.
We pa\-ra\-meterize $\mathcal{G}_{\theta}$ as a linear feedforward neural network, with $\theta$ being the learnable parameters. 
Unlike differential equations, our eigenvector problem does not lend itself to a Fourier-based structure, and thus we adopt such a simpler architecture. As we will demonstrate, this straightforward model performs effectively across various quantum systems and different correlation regimes. 

For the sake of clarity, Fig.~\ref{fig:learning} illustrates the differences between the learning schemes for the unitary part of the exact many-body ansatz, i.e., the mapping between the set of parameters ${\bm{f}, \bm{A}^{(1)}, ..., \bm{A}^{(n-1)}}$ and $\bm{A}^{(n)}$. In the standard contracted quantum eigensolver approach, each set $\bm{A}^{(n)}$ is learned sequentially from the previous ansatz parameters $\{\bm{A}^{(m)}\}_{m=1}^{n-1}$ and the Hamiltonian parameters $\bm{f}$ (dotted blue lines). In contrast, the surrogate neural network solver proposed in this work directly learns all the sets $\{\bm{A}^{(m)}\}_{m=1}^{n}$ from $\bm{f}$ (solid orange lines), extending the scheme to the entire set of learnable ansatz parameters i.e., $n \mapsto k$.

\section{Results}

The first system for which we build a neural solver is the
generic $M$-qubit Hamiltonian:
\begin{align}
    \hat H = \sum^{3}_{r_1 =0} \cdots \sum^{3}_{r_M =0} f_{r_1,...,r_M} \sigma_{r_1} \otimes \cdots  \otimes\sigma_{r_M}   \,,
    \label{hamiltonian}
\end{align}
where $\sigma_r$ denotes the Pauli matrix. In this system, the number of terms of the Hamiltonian scales as $4^M$. The Hamiltonian contains the entire set of generators of the Lie algebra of the corresponding unitary group and thus, in this particular example, is possible to solve the eigensystem with $k =1$ and the unitary part suffices. We let all the Hamiltonian parameters vary in given regimes and solve the problem by finding the ansatz with a classical contracted eigensolver. We then use those data to train a neural network and this neural network should be able to predict the ansatz parameters from the ones of the input Hamiltonian. 

\begin{figure}[t]
\includegraphics[width = \linewidth]{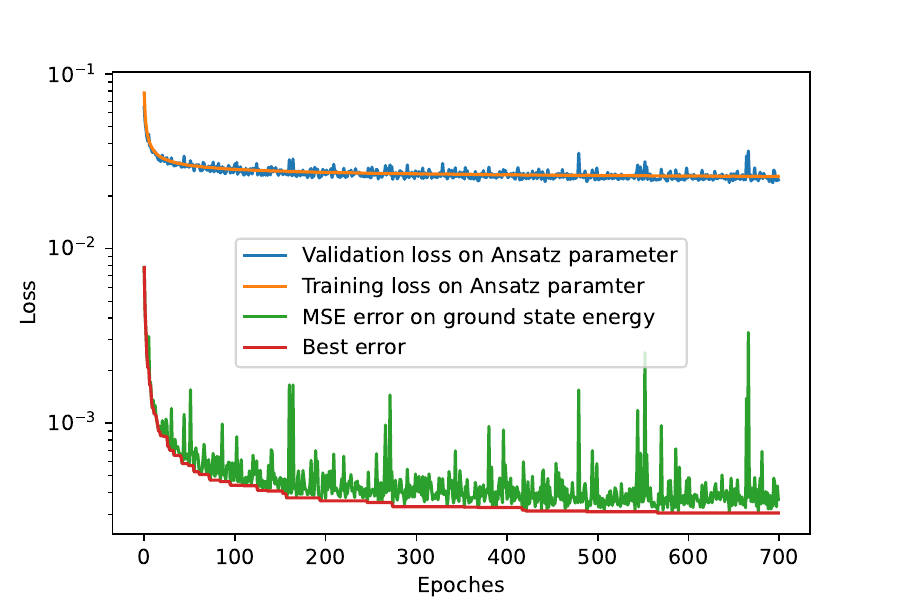}
\caption{Validation loss and MSE on the ansatz parameters and ground state energy for the Hamiltonian \eqref{hamiltonian}, in the parameter regimen $f_{r_1,r_2} \in (-0.2,0.2)$. The final validation loss is 0.0247, and the final error on ground state energy is 0.000305, at epoch 700. 
}
\label{fig:training1}
\end{figure}

\begin{figure}[t]
\includegraphics[width = \linewidth]{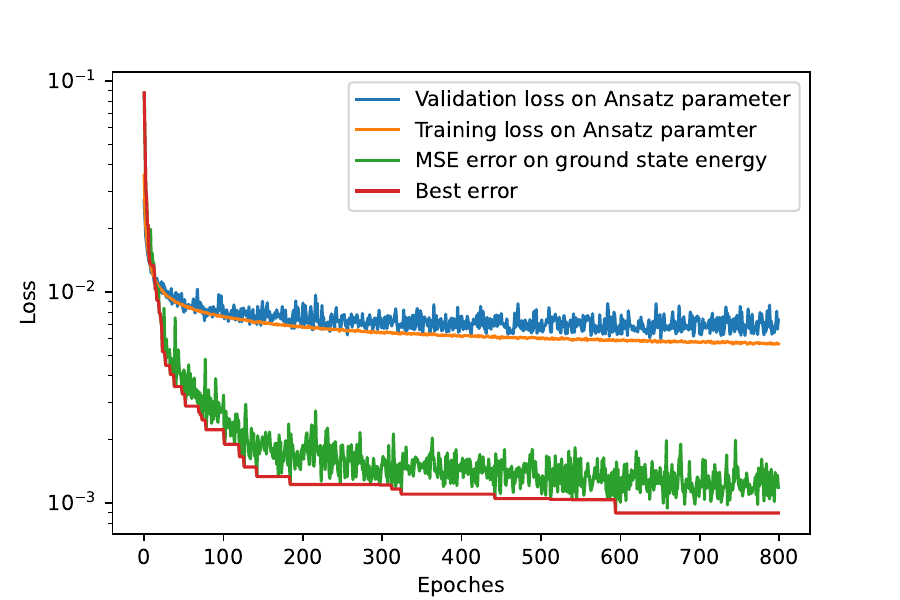}
\caption{Validation loss and MSE on the ansatz parameters and ground state energy for the Hamiltonian \eqref{hamiltonian}, in the parameter regimen $f_{r_1,r_2} \in (-3.8,-1.2)$. The final validation loss is 0.00735, and the final error on ground state energy is 0.000894, at epoch 800. 
}
\label{fig:training2}
\end{figure}

We present the results for the case of $M = 2$, where the neural network model inputs 16 Hamiltonian parameters and outputs 16 ansatz parameters. Our architecture contains 6 hidden layers of dimension 256. To improve the stability of the model, all hidden layers are residual layers. The loss function is the mean-squared error (MSE) loss between predicted ansatz parameters and precomputed data. The learning rate is 0.0001 for the Adam optimizer. The training is accomplished by a single NVIDIA A40/A100 GPU. We trained the model by changing all the parameters $f_{r_1,r_2}$ for two different regimes (as we will explain later, these regimes were chosen to avoid energy crossings). The first regime $f_{r_1,r_2} \in (-0.2,0.2)$ and the second $f_{r_1,r_2} \in (-3.8,-1.2)$. This is undoubtedly a vast parameter space. Ref.~\cite{ceroni2023generating} reported that three data points were sufficient to train the UCC ansatz with single and double excitations for the dissociation of the hydrogen molecule with minimal basis set. Based on this result, we estimate that three data points per parameter are needed to train our more robust ansatz, requiring a total of $3^{16}$ data points. However, we found that much fewer data points were needed for training. We observed convergence in both regimes with just 20,000 data points. Our results using 4 million data points, an order of magnitude fewer than our estimate, are shown in Fig~\ref{fig:training1} and Fig.~\ref{fig:training2}. The final validation loss for the ansatz parameters is 0.02 and 0.007, while the error in the ground state energy is $3 \times 10^{-4}$ and $8 \times 10^{-4}$, respectively. We also test various neural network architectures, including a convolutional layer with 2-dimensional input data $f_{r_1,r_2}$ \cite{ZHOU202219}, and various loss functions, including MSE loss and relative loss, or a mix of them. Interestingly, we found no appreciable improvement in the model's performance. We attribute this behavior to the robustness of the exponential ansatz. 

\begin{figure}
\begin{center}
\includegraphics[width = \linewidth]{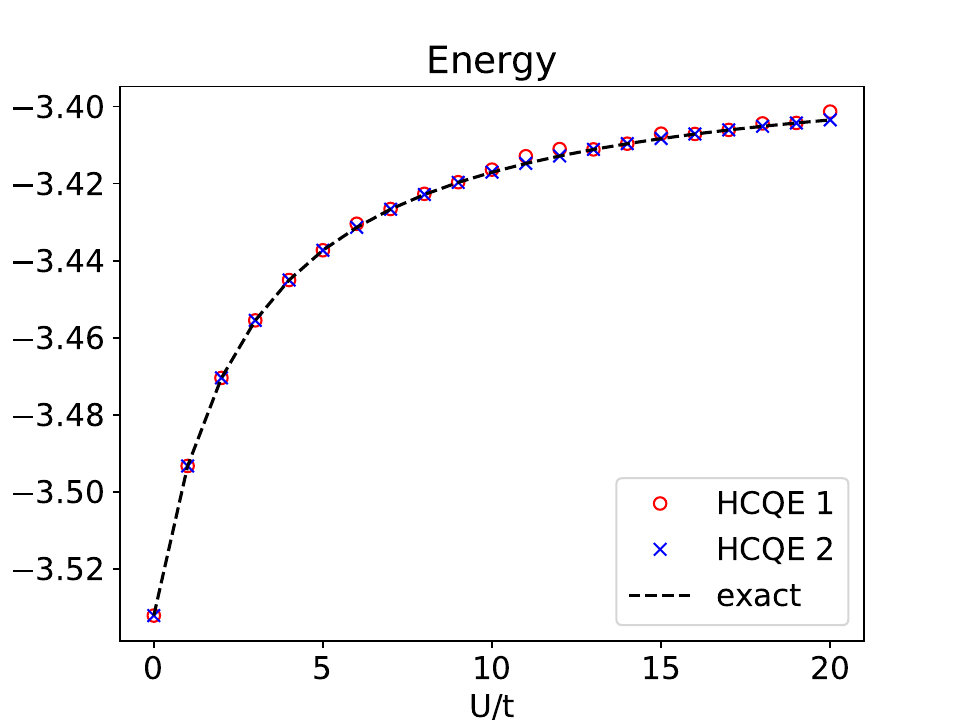} 
\end{center}
\begin{center}
\includegraphics[width = \linewidth]{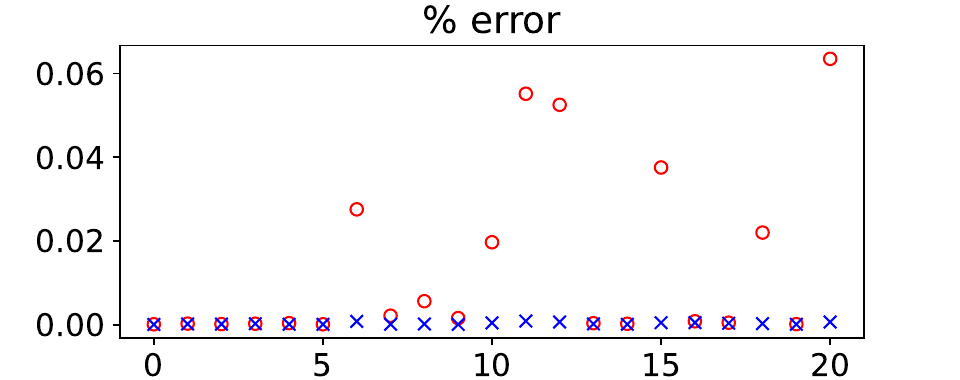}
\caption{Upper panel: The exact and predicted ground-state energies for two iterations of the 2-CQE method are shown as a function of the relative interaction strength, \( U/t \), for the Fermi-Hubbard Hamiltonian in Eq.~\eqref{hamiltonianfermi} with \( L = 9 \) sites and \( N = 2 \) particles. By the second iteration, the method successfully captures the exact ground-state energy. Lower panel: For further illustration, we present the percentage error in the energy for both iterations.}
\label{fig:HCQE}
\end{center}
\end{figure}

The presented results clearly show the possibility of training a neural network that learns the implicit functional between the Hamiltonian and the ansatz parameters. Still, our results can be influenced by the low dimensionality of the Hilbert space of the model. To challenge our neural network we consider a 1D lattice of interacting fully polarized fermions. The system is described by the following Hamitonian:
\begin{align}
\label{hamiltonianfermi}
    \hat H = - t\sum^L_{m=1} (\hat c^\dagger_m \hat c_{m+1} + \hat c^\dagger_{m+1} \hat c_{m}) + U \sum^L_{m=1} \hat n_m \hat n_{m+1}\,.
\end{align}
Here the operator $\hat c^\dagger_m$ ($\hat c_m$) creates (annihilates) a fermion on lattice site $m$, and $U$ is the strength of the nearest-neighbor repulsion. This Hamiltonian plays an important role in the discovery of disorder-free many-body localization \cite{vanNieuwenburg9269,PhysRevLett.122.040606}. We impose periodic boundary conditions. To study the performance of the non-unitary part of the ansatz, we disregard the unitary terms. In the literature, this is called the Hermitian CQE (HCQE) \cite{Smart.2024}. Quite remarkably the HCQE recovers the exact ground state in only 2 iterations, i.e., $k =2$ in Eq.~\eqref{ansatzfinal}, using as numerical solver the limited-memory quasi-Newton L-BFGS method \cite{Liu1989}. The trial wave function is the ground state at $U=0$. As an example, in Fig.~\ref{fig:HCQE} we show the exact ground-state energy and the results of the 2-CQE highlighting the energies of the first (HCQE 1) and the second (HCQE 2) iterations for $L = 9$ sites and $N = 2$ particles with the interacting range of $U/t\in[0,20]$. The second iteration captures the exact ground state of the system in this fairly large range of quantum correlations. The first iteration has already performed quite well with an average energy error of $0.013\%$. The average error after the second iteration is $0.0003\%$. 

\begin{figure}[t]
\includegraphics[width = \linewidth]{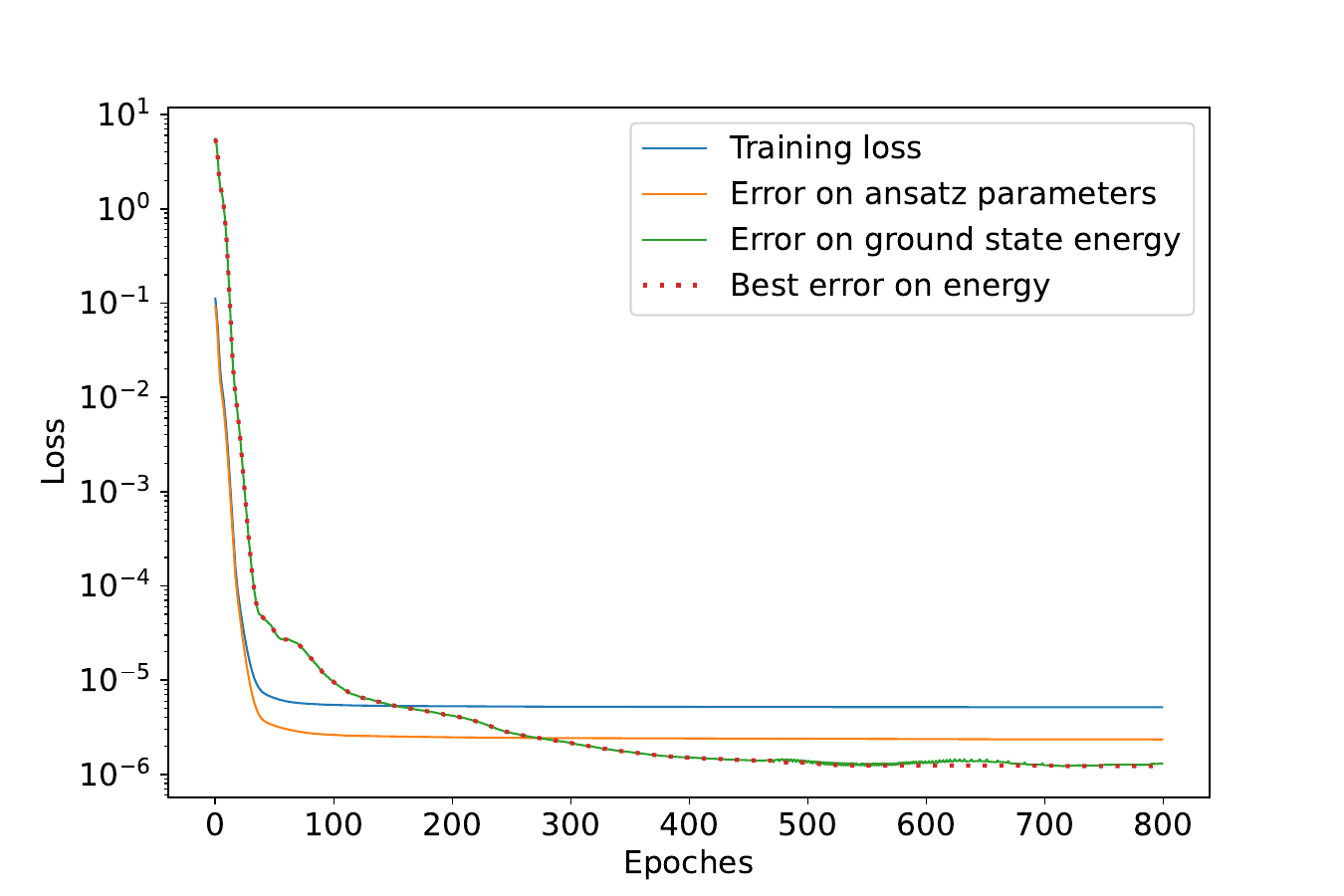}
\caption{Training loss, error on the ansatz parameters and error on the ground state energy for the Fermi-Hubbard Hamiltonian \eqref{hamiltonianfermi} with $L=5$, $N = 2$, and $U\in (0,20)$.}
\label{fig:training-hubbbard-L5}
\end{figure}

\begin{figure}[t]
\includegraphics[width = \linewidth]{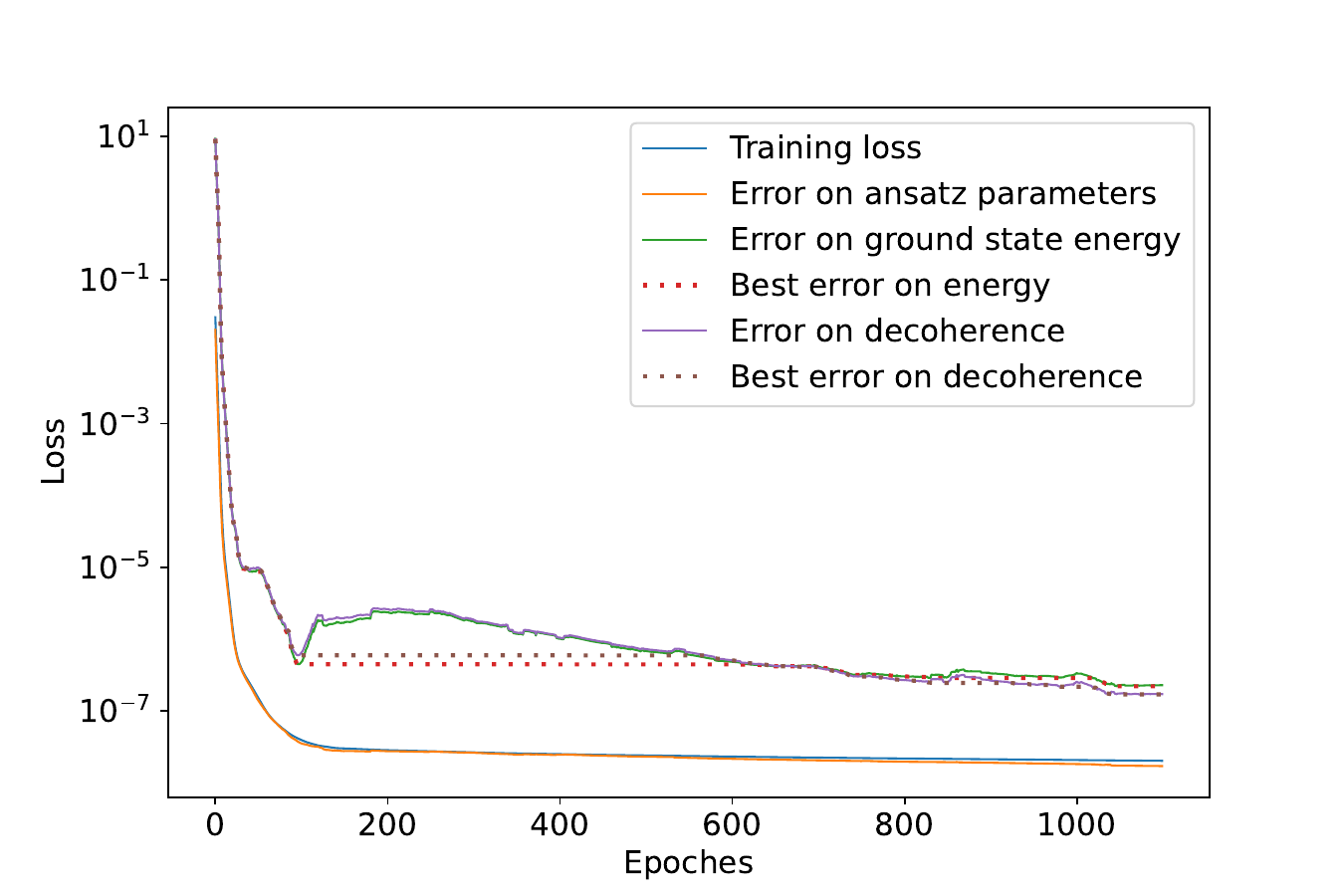}
\caption{Training loss, error on the ansatz parameters and error on the ground state energy and decoherence for the Fermi-Hubbard Hamiltonian \eqref{hamiltonianfermi} with $L=8$, $N = 2$, and $U\in (0,20)$.}
\label{fig:training-hubbbard-L8}
\end{figure}

We now discuss the results of the neural-network exponential ansatz for the Fermi-Hubbard model. Notice that at each iteration the ansatz contains $2L$ parameters. Since only two iterations are needed in this case, we learn $4L$ parameters, preventing the exponential growth of the more popular generalized UCC. Fig.~\ref{fig:training-hubbbard-L5} and Fig.~\ref{fig:training-hubbbard-L8} present the results obtained by the neural-network solver for $L=5$ and $L=8$ sites with 2 fermions. The architecture is similar to the one used for the $M$-qubit Hamiltonian. We sample 2000 data points in the same regime of interaction depicted in Fig.~\eqref{fig:HCQE}. We observe a quite fast convergence for the ansatz parameters and a high accuracy for the ground-state energy. We also studied the behavior of a more interesting observable, the non-diagonal part of the one-body reduced density matrix $\langle \hat c^\dagger_{m+1} \hat c_m\rangle$, which exhibits a similar convergence as the energy.    

Our numerical simulations show that the neural network is able to learn the entire set of ansatz parameters of quantum many-body systems in certain interacting regimes. It is worth mentioning that to cover all parameter regimes, one has to deal with energy crossing or conical intersections. Yet, the neural network itself is in principle a large fitting function to mimic a smooth data map, where small changes in input should only cause small changes in output as well. As a result, our neural network alone is not enough to describe a jump in the ansatz parameters, which happens when the symmetries of the states change abruptly as a result of an energy crossing or a conical intersection. To deal with such jumps in the context of potential energy surfaces, some methods eliminate a small vicinity to cover both sides of the conical intersection~\cite{Wang2023}. While we left for future work to train a neural solver for the entire set of parameters, we observe that upon a conical intersection, the ground state changes to another state that has a lower energy. Instead of training a map only for the ground state, one can train for multiple states and use fast post-processing to filter the ground state with the lowest energy. As a second solution, although a single neural network cannot converge to a jump function, it can describe functions with discrete input ranges, for example, all odd-output regimes of a math floor function. Then one can use two neural networks to describe the odd-output and even-output regimes respectively, thus covering the entire regime of the math floor function.


\section{Conclusions} 

Inspired by recent progress in operator learning techniques, we introduced a neural network universal exponential ansatz for quantum many-body physics. Our surrogate model $\mathcal{G}_{\theta}: \bm{f} \rightarrow \{\bm{A}^{(n)},\bm{B}^{(n)}\}$ captures the implicit functional relationship between the Hamiltonian and the parameters of the universal, formally exact exponential ansatz used on modern contracted quantum eigensolvers.  Despite the complex nonlinear recursion governing the ansatz parameters, which resists a compact formulation, a linear feedforward neural network achieved high accuracy across various quantum systems and correlation regimes. Quite remarkably, once trained, our model not only removes the need for repetitive and computationally intensive calculations but also reinforces the universality of this approach. In stark contrast to the more common neural approach based on the amplitudes of the wavefunction \cite{Lange_2024}, our operator-learning-inspired approach to exact exponential ans\"atze offers a novel and powerful approach to tackling the quantum many-body problem, especially in correlation regimes that are beyond the reach of traditional (functional, coupled-cluster, or mean-field) theories. 

Finally, since the ansatz studied in this work is connected with a growing family of quantum eigensolvers, we believe that our results can be used as a pathway for extracting quantum circuit parameters for state preparation from the neural network ansatz parameters, which we leave for future work. We have not yet discussed the choice of basis for expressing the second-quantized Ha\-mil\-tonian, which plays a crucial role in quantum chemistry and material science \cite{doi:10.1021/acs.jctc.6b00156,doi:10.1021/acs.jctc.4c00528,langkabel2024adventfullyvariationalquantum,doi:10.1021/acs.jpclett.3c02536}. Future work will also explore the role of optimized basis sets in our surrogate model.

\begin{acknowledgments}
W.Z.~would like to thank helpful discussions with Xiangping Jiang and Shengliang Yu, and the support of the National Science Foundation of China (Grant No. 62301505). W.Z., J.Z. and L.P.C.~acknowledge support via a starting grant of the research center of new materials computing of Zhejiang Lab (3700-32601). C.L.B.-R.~thanks insightful discussions with David Mazziotti, Alán Aspuru-Gu\-zik, and Artur Izmaylov, and gratefully acknowledges financial support from the Royal Society of Chemistry and the European Union’s Horizon Europe Re\-search and Innovation program  un\-der the Marie Skło\-dowska-Curie Grant Agreement n°101065295. Views and opinions expressed are however those of the author only and do not necessarily reflect those of the European Union or the European Re\-search Executive Agency.
\end{acknowledgments}

\section*{Data availability statement}

All codes and data to reproduce, examine, and improve our proposed analysis are freely available online at the following URL: \url{https://github.com/cabenav/ACSE-universal-ansatze}.

\bibliography{mixqcse}


\end{document}